\documentclass{article}

\usepackage{PRIMEarxiv}

\usepackage[utf8]{inputenc} 
\usepackage[T1]{fontenc}    
\usepackage{hyperref}       
\usepackage{url}            
\usepackage{booktabs}       
\usepackage{amsfonts}       
\usepackage{amsmath}
\usepackage{nicefrac}       
\usepackage{microtype}      
\usepackage{lipsum}
\usepackage{fancyhdr}       
\usepackage{graphicx}       
\usepackage{natbib}
\usepackage{eurosym}
\usepackage{booktabs,tabularx}
\usepackage{rotating}
\usepackage{pdflscape}
\usepackage{graphicx}
\usepackage{subcaption}

\title{FleetPy: A Modular Open-Source Simulation Tool for Mobility On-Demand Services
}

\author{
  Roman Engelhardt, Florian Dandl, Arslan-Ali Syed, \\ \bf{Yunfei Zhang, Fabian Fehn, Fynn Wolf and Klaus Bogenberger} \\
  Chair of Traffic Engineering and Control \\
  Technical University of Munich \\
  Arcisstraße 21 \\
  80333 Munich, Germany \\
  \texttt{Corresponding Author: roman.engelhardt@tum.de} \\
}

\begin{document}
\maketitle

\begin{abstract}
The market share of mobility on-demand (MoD) services strongly increased in recent years and is expected to rise even higher once vehicle automation is fully available. These services might reduce space consumption in cities as fewer parking spaces are required if private vehicle trips are replaced. If rides are shared additionally, occupancy related traffic efficiency is increased. Simulations help to identify the actual impact of MoD on a traffic system, evaluate new control algorithms for improved service efficiency and develop guidelines for regulatory measures. This paper presents the open-source agent-based simulation framework FleetPy. FleetPy (written in the programming language "Python") is explicitly developed to model MoD services in a high level of detail. It specially focuses on the modeling of interactions of users with operators while its flexibility allows the integration and embedding of multiple operators in the overall transportation system. Its modular structure ensures the transferabillity of previously developed elements and the selection of an appropriate level of modeling detail. This paper compares existing simulation frameworks for MoD services and highlights exclusive features of FleetPy. The upper level simulation flows are presented, followed by required input data for the simulation and the output data FleetPy produces. Additionally, the modules within FleetPy and high-level descriptions of current implementations are provided. Finally, an example showcase for Manhattan, NYC provides insights into the impacts of different modules for simulation flow, fleet optimization, traveler behavior and network representation.
\end{abstract}


\section{Introduction}
The mobility landscape is probably about to enter a phase of major change, enforced by transportation related problems, like crowded or congested infrastructure and related local and global emissions. On the demand side, urbanization leads to growing cities~\cite{UN2022} and the increasing number of trips causes considerable problems, which are becoming more and more apparent~\cite{Rodrigue2020}. Due to limited space, the infrastructure cannot grow at the same rate as the demand, especially for motorized traffic. Hence, congestion during peak hours continues to grow. Besides increasing delays for travelers, this also worsens the already existing local pollutant and noise emissions, thereby impacting the health of residents directly. Furthermore, the world has to deal with globally increasing greenhouse gas emissions~\cite{EEA2022}.

Technological advances, namely digitalization and the mobile internet, enabled the rise of new mobility-on-demand (MoD) services, such as Didi Chuxing, Uber, Lyft, ViaVan, MOIA, and many others. These services try to enrich the transportation offer, in addition to the established means of transportation. Topics such as first/last-mile mobility solutions, which aim at increasing the accessibility of public transportation, or integrated transportation approaches, combining different forms of mobility, are also becoming more popular. MoD services are usually user-centric and highly convenient and bridge the gap between private and public transportation: on the one hand, they can be viewed as part of the public transportation system as private vehicle ownership is no longer necessary; on the other hand, the service offers are in most cases specific to users and the capacities are more reminiscent of private transportation.

Hand in hand with the development of new forms of mobility services, social trends have a significant influence on people's mobility behavior. In recent years, a trend toward a more sustainable lifestyle could be observed among broad segments of the population. This also includes the phenomenon of no longer wanting to own things, but to share them with others instead, namely sharing economy~\cite{heinrichs2013sharing, standing2019implications}. MoD services, especially so-called transportation network companies (TNCs), became very popular, taking over significant mode shares in several cities around the globe~\cite{tirachini2020ride, acheampong2020mobility}. The relevance of these services is expected to grow even further with the introduction of autonomous vehicles, which will provide better cost structures and new operator models~\cite{Bosch.2018,kuhnimhof2021mobility,Dandl.2019b}.

Technological advances and user-centric services are no guarantee for a better future for the community. For instance, extensive use of ride-hailing services, allowing only one request per vehicle, is likely to increase congestion~\cite{schaller2021can}. However, ride-pooling can alleviate the stress on the street network by serving multiple passengers with the same vehicle~\cite{Engelhardt.2019}. In this context, it is of high interest to study and introduce regulatory measures for the emerging mobility services in order to enrich the overall mobility offer of a city~\cite{Dandl.2021}. To identify these measures and provide recommendations to the regulators, the simulation and evaluation of MoD service models are of great economic and social interests.

This paper presents the open-source simulation framework Fleetpy\footnote{https://github.com/TUM-VT/FleetPy}, which is designed to study these MoD services in detail. The subsequent sections provide an overview of existing simulation tools and reasons why we decided to develop FleetPy. These are followed by a high-level software description and its required data to build a simulation model. This paper includes a reference show case highlighting some of FleetPy's capabilities. Finally, the paper is concluded by a summary and future research ideas.

\subsection{Literature Review}

There are several simulation frameworks to model vehicle fleets. They provide different focuses, and thus have different strengths and weaknesses. In their review papers, \citep{Nguyen2021} and \citep{Huang2022} give an overview of agent-based traffic simulators and elaborate how and where these simulators can be used. Their studies discuss not only the technical aspects of the simulators but also present the different system components required for each simulation framework. The focus of the following literature review lies on frameworks specializing on the representation of MoD services. In addition to licensing and programming language, the frameworks are classified by investigating their representation of vehicle and network aspects (e.g. representation of traffic flow, travel times and stops), demand modeling and user-operator interaction (e.g. activity-based or exogenous demand and operator and user choice modeling) and additional fleet control modules (e.g. (re-)assignment, repositioning, pricing, charging, etc.), including real-world case studies.

Among the commercial software, the solutions of Aimsun and PTV are particularly worth mentioning. For the simulation of MoD fleets, Aimsun presents "Aimsun Ride" \cite{AimsunRide2022} as a fleet management software for Mobility as a Service (MaaS). According to their documentation, it can represent virtual MoD stops or a door-to-door service. Furthermore, different fleet types and configurations are implemented while vehicles can be dispatched either based on fixed timetables or dynamically for on-demand or pre-booked requests. In addition, the user can choose between different pricing and charging models, or implement competition between different mobility providers. The PTV Maas Modeller \cite{PTVMaaSModeller2022} offers a REST API. It allows to calculate mode choices and incorporate a demand model from regional planning (VISUM). Thereby, the user can model impacts on travel times, accessibility to other transportation offers and the cost of operation.

Among the open-source simulators that include MoD solutions, MATSim and SUMO likely have the largest user bases, however, other simulators like SimMobility, Polaris or MaaSSim also offer interesting solutions. MATSim \cite{Horni2016} is an agent-based transportation simulation framework with many extensions. It has a so-called DRT module and an extension called AMODEUS \cite{Ruch2018} to model on-demand fleets. In an iterative simulation process, the bookings from an activity-based pre-day demand model are served by a mobility fleet in the subsequent transportation simulation, which are then scored before the next iteration. SimMobility \cite{Adnan2016} is also an agent-based model that includes scheduling-, mode- and route- choice. Agents can choose destination and departure time on a multi-modal transportation network in an activity based demand model. It offers a three-level modeling approach (i.e. long-term, mid-term and short-term) and integrates an interface for agent-to-agent communication. The microscopic simulation software SUMO in combination with Jade \cite{Soares2014} uses activity diaries, which are input for the fleet management performed in Jade. The simulation is based on a SUMO road network, on which microscopic vehicle movements are displayed. Polaris \cite{Auld2016} integrates a travel demand model as an input for an agent-based simulation. It is fully integrated, activity-based and contains dynamic network travel times. The network and fleet operation simulation is performed in a single process and can, according to the authors, be rapidly adapted to new emerging applications. MaaSSim \cite{Kucharski2022} is an agent-based simulator for two-sided mobility platforms, in which the interactions of the operator with both travelers and drivers is modeled. The model consists of a mobility platform matching demand and supply and relies on an urban road network graph, travel demand and supply specifications as inputs.

An overview of the mentioned simulators can be found in Table~\ref{tab:fleet_simluation_frameworks}, describing the respective network, demand modeling, and fleet control modules. The MATSim and FleetPy frameworks are capable to model multiple operators. The column "Network Modeling" refers to the travel times of fleet vehicles and MoD stop modeling. The "Demand Modeling" section refers to the passenger demand generation and the user-operator interaction of the respective simulators. Here, the modeling of the passenger decisions is of special interest, as real-world app usage with real-time information can only be modeled by within-day choice models by incorporating user requests and operator decision/offers. The "Fleet Control Modules" represent additional functionalities and also give a first impression of the modularity and extendability of the frameworks. It can be seen that the additional functionalities of FleetPy are an advantage compared to most simulators. With FleetPy one can easily create new control modules for specific studies or use old ones as a basis for new developments.

\begin{landscape}
\begin{table}[ht]
\caption{Overview of existing open-source fleet simulation frameworks}
\label{tab:fleet_simluation_frameworks}
\begin{tabular}{p{4cm}|p{2.5cm}|p{2cm}p{1cm}|p{2cm}p{2.5cm}|p{5cm}}
\toprule
\textbf{Name} & \textbf{Programming} & \multicolumn{2}{c}{\textbf{Network Modeling}} & \multicolumn{2}{c}{\textbf{Demand Modeling}} & \textbf{Fleet Control} \\
~ & \textbf{Language} & Travel Times & Stops & Generation & User-Operator Interaction & \textbf{Modules}\\
\midrule
MATSim \cite{Horni2016} and AMoDeus \cite{Ruch2018} & Java & dyn:meso & vis, d-t-d & abm & u:b & u:re, repo, pr, ch, log \\
\midrule
SUMO and JADE \cite{Soares2014} & Java & dyn:micro & d-t-d & ex & u:b & \\
\midrule
SimMobility \cite{Adnan2016} & C++ & dyn:meso & d-t-d & abm & u:b, u:r $\to$ o:d & pr, log \\
\midrule
POLARIS \cite{Auld2016} & C++ & dyn:meso & d-t-d & abm & u:b &  \\
\midrule
MaaSSim \cite{Kucharski2022} & Python & det:con & d-t-d & ex & u:r $\to$ o:d, u:r $\to$ o:o $\to$ u:d & repo, pr \\
\midrule
FleetPy & Python & det:tv, dyn:meso & vis, d-t-d & ex & u:r $\to$ o:d, u:r $\to$ o:o $\to$ u:d & u:re, repo, pr, ch, log \\
\bottomrule
\end{tabular}
\newline
\underline{Abbreviations:}
\newline
\textit{Traffic Flow and Travel Times:} \textbf{dyn}amic \textbf{meso}scopic traffic flow model, \textbf{dyn}amic \textbf{micro}scopic traffic flow model, \textbf{det}erministic \textbf{con}stant, \textbf{det}erministic \textbf{t}ime \textbf{v}arying
\newline
\textit{Stop Representation:} \textbf{d}oor-\textbf{t}o-\textbf{d}oor, \textbf{vi}rtual \textbf{s}tops
\newline
\textit{Demand Generation:} \textbf{a}ctivity \textbf{b}ased \textbf{m}odel, \textbf{ex}ogeneous request set
\newline
\textit{User-Operator Interaction:} \textbf{u}ser \textbf{b}ooking, \textbf{u}ser \textbf{r}equest $\to$ \textbf{o}perator \textbf{d}ecision, \textbf{u}ser \textbf{r}equest $\to$ \textbf{o}perator \textbf{o}ffer $\to$ \textbf{u}ser \textbf{d}ecision
\newline
\textit{Additional Modules:} \textbf{u}ser \textbf{re}-assignment, \textbf{repo}sitioning, \textbf{pr}icing strategy, \textbf{ch}arging strategy, \textbf{log}istic schemes
\end{table}
\end{landscape}

\subsection{FleetPy Features}
As shown in the previous section, several simulation frameworks can model MoD services. They were mostly created to answer research questions from an operational viewpoint or to study the role of MoD services within the entire transportation system of a city. In the surveyed simulators, the \textbf{real-world interactions between users and MoD operator} are not modeled with realistic and easily extendable parameters, i.e. the user receiving real-time information from a mobile phone application (app) and making decisions based on the provided information. Moreover, simulation environments often lack the \textbf{flexibility} to integrate multiple MoD providers using digital platforms with real-time information. This facilitates the realistic modeling of systems like a broker-based system, an integrated system of MoD services and public transportation network, or a combined system of MoD passengers and freight.

Finally, FleetPy is built as a modular framework that enables the \textbf{transferability} of prior developments to new strategies and selection of \textbf{appropriate level of detail}. For example, algorithms implemented for assigning customers to vehicles can be directly applied, without any changes, to study a newly developed fleet repositioning algorithm. Selecting the appropriate level of detail can be best explained by giving two examples: 1) for specific operational studies, an analyst might prioritize computational speed over a detailed representation of city-wide network dynamics and 2) a study on regulatory requirements with many simulation scenarios might prefer an acceptable run-time for MoD operational algorithms and invest computational resources on modeling better network dynamics. The level of details provided via FleetPy modules and possible parameters enable FleetPy users to adopt the simulation according to their requirements. In summary, FleetPy provides a hierarchical framework where straightforward fleet control algorithms and models can be exchanged with more elaborate solutions by changing simulation modules and parameters. 

\section{FleetPy Software Description}\label{sec:software_description}

\begin{figure}[tbp]
    \centering
    \includegraphics[width=0.9\textwidth]{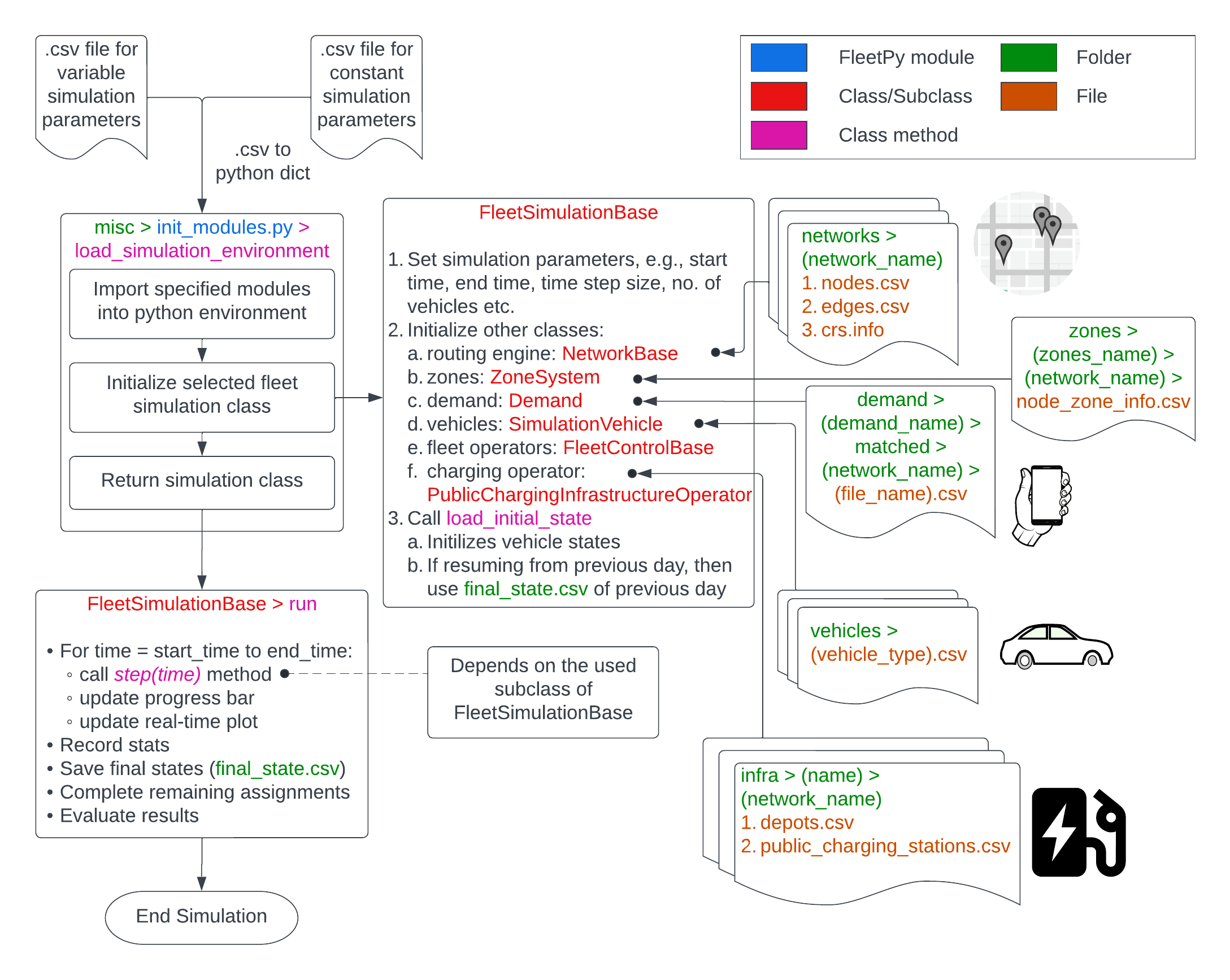}
    \caption{Main simulation flow of FleetPy}
    \label{fig: main_loop_flow_chart}
\end{figure}

The main structure of simulating FleetPy scenarios is illustrated in Fig.~\ref{fig: main_loop_flow_chart}. The required input data can be divided into two categories: input parameters and data files. 
Input parameters describe the FleetPy modules and configuration to use for the simulation. Scenario definition files handle these input parameters. Since FleetPy relies on several parameters, while typically only a few are varied across multiple simulation runs, two input files are used. The first contains all constant parameters, while the second consists of the parameters that vary among simulations. This helps an analyst to clearly view the differences between distinct simulation scenarios. 
While the input parameters control which modules and parts of the code are executed, data files provide structured inputs. These are necessary to describe e.g. networks, zones, customer demand, vehicles and infrastructure. Data files are loaded along with the modules in the initialization of the simulation (\textit{FleetSimulationBase}, the parent class for the core simulation). The simulation modules selected in the input parameters determine which code is executed during the \textit{step} method of the simulation flow as shown in Fig.~\ref{fig: main_loop_flow_chart}. The simulation creates disaggregated records for vehicles, users, and --- if applicable --- the infrastructure. Since an analyst is usually more interested in the bigger picture, automatic evaluation functions generate aggregated outputs from these records. Detailed requirements for the input data are listed in the next section.


\subsection{FleetPy Input and Output Data}\label{subsec:input_output_data}
Files that provide structured data --- usually in csv-format --- describe both supply and demand.

The network is represented as a routable directed graph with edges as connection between nodes. The minimal network input for FleetPy must contain node and edge information. A nodes-file provides information about node-id and geographical location while an edges-file defines connections by start- and end-node-ids and edge travel times and distances. Time-dependent travel times can be added by either an additional file specifying aggregated travel time factors or updated edge travel times for given simulation times.

Travelers are loaded in the demand module via a csv-file that lists all travelers along with their request id, request time, and start and end locations. The locations refer to node indices and therefore the travelers must be matched to the applied network. Additional traveler specific attributes defining mode choice and dynamic behavior can be added.

Multiple vehicle types can be initialized for the MoD fleet. The parameters defining the vehicle type are read from a csv-file and specify vehicle capacity (how many passengers this vehicle can carry), costs (both fixed costs and distance-based costs), battery size and range for electric vehicles.

Optionally, vehicle infrastructure can be added to model the interaction of vehicles with specific network locations providing special infrastructure for the MoD vehicles. This infrastructure can define depots (charging and parking infrastructure dedicated to a distinct operator), public charging stations (charging infrastructure available for all vehicles) or designated pick-up and drop-off points for the MoD service(s)~\cite{Engelhardt.2021}. Input-files for each infrastructure type provide information about network location and, depending on the type of infrastructure, number of parking spots and/or number of charging slots and their power.

For some use cases (i.e. forecasting and repositioning or network clustering and macroscopic fundamental diagram (MFD) estimation to scale network travel times~\cite{Dandl.2021}) the aggregation of the network into zones is required. FleetPy supports defining zones within the road network which are optionally imported using the zone description file. The zone data maps each node to a zone along with zone geometry and centroid nodes which are used for routing on zone level. Furthermore, preprocessed demand forecast files can be imported that provide estimations for future trips within certain zones and serve as information source for the fleet control's repositioning module.



Once a simulation starts, a result folder is generated and output of the simulation is continuously written into output files within the folder. Outputs include:

\begin{itemize}
    \item User statistics: information about each traveler's trip request (acceptance/rejection, pick-up/drop-off time and location, offer parameters and operator, service vehicle)
    \item Operator statistics: the completed information (time, vehicle ID, location, and order) of each individual task such as a pick-up or drop-off journey, boarding, waiting, etc.
    \item Time-dependent statistics: attributes evaluated in each simulation time step (i.e. computational time)
    \item Log-files: the detail of log messages can be controlled for bug-fixing purposes
\end{itemize}

Other outputs vary from research focuses such as charging activities or repositioning records. At the end of the simulation, final states of all fleet vehicles are recorded. Additionally, a standardized evaluation computes key performance indicators (KPIs) for the finalized simulation.

\subsection{Core Simulation Modules}
\begin{figure}
    \centering
    \includegraphics[width=\textwidth]{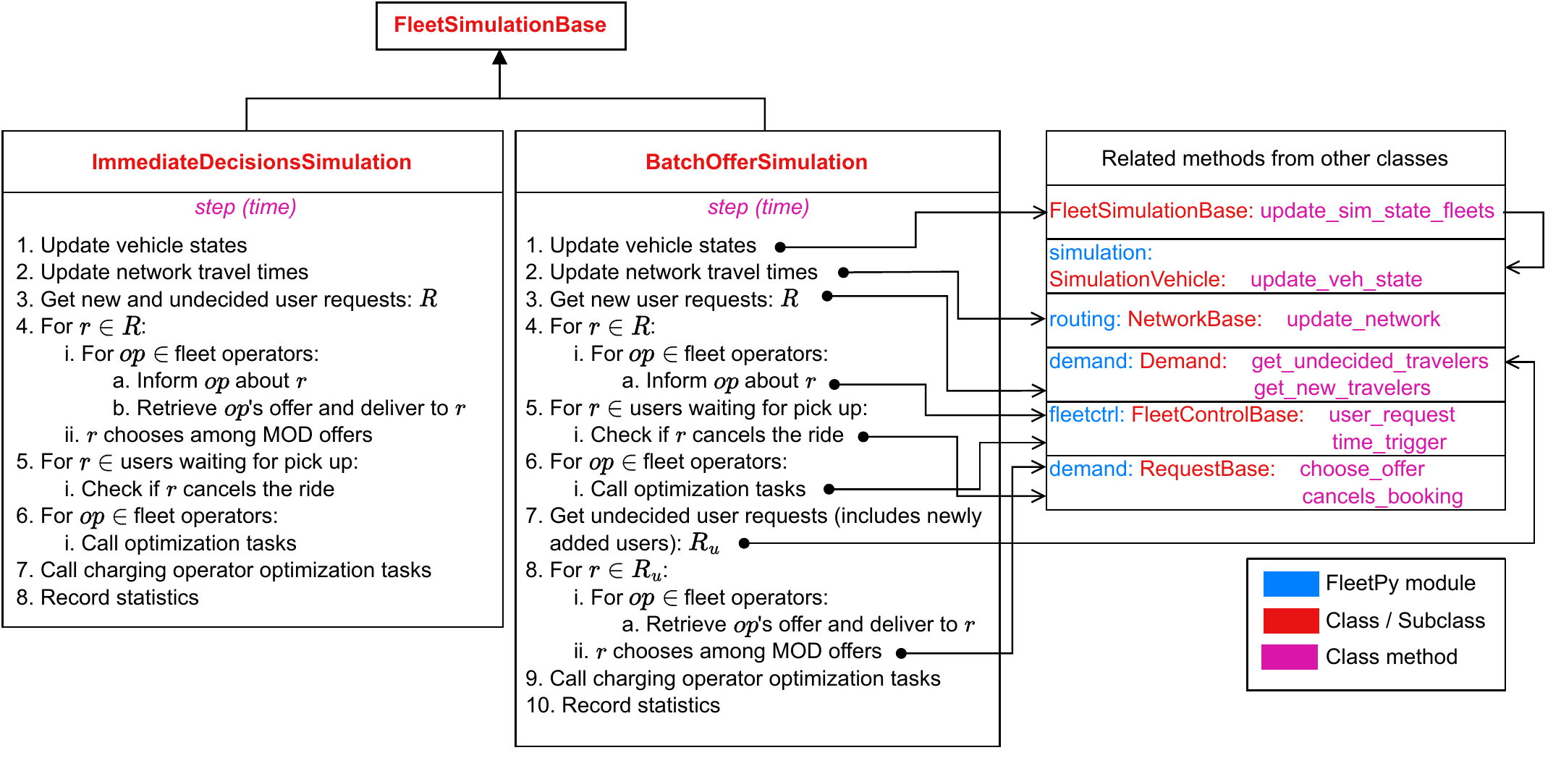}
    \caption{Implementations of the step method of the simulation class}
    \label{fig: subclasses flow chart}
\end{figure}

After initializing both the core modules and the data for network, infrastructure, vehicles, and demand, the framework loops over simulation time $t$. Currently, there are two main implementations of this time loop, which are depicted in Fig.~\ref{fig: subclasses flow chart}, denoted by ImmediateDecisionsSimulation (IDS) and BatchOfferSimulation (BOS). The key difference between both is the interaction model between MoD customers (travelers that request a trip for an MoD operator) and operators. 

In the IDS, the interaction between customer and MoD operator consists of three parts, which are performed immediately one after the other. The customer requests service from the MoD operator, which, as a response, triggers either an offer or a rejection by the operator. This response is created by the fleet control module in the method \textit{user\_request}. If an offer is sent, the customer has the option to book the MoD service or cancel the request. The decision is modeled in the demand module as customer behavior (method \textit{choose\_offer}). In the IDS, the customer decides and communicates the decision to the operator immediately. Subsequently, the next customer request is handled.

The interaction is more complex in the BOS. Here, the customer requests are only registered by the operator. Instead of answering each request immediately, the operator batches open requests for a specified amount of time and creates offers for all unanswered requests simultaneously. In a separate step, the customers with open requests make decisions. In this framework customers must not decide immediately for or against an offer. Instead, a waiting time for other operators to answer or dynamic decision making of customers can be modeled. 

Both frameworks have a state update phase at the beginning of each time step and a time-triggered fleet control phase after all new requests are received. In the update phase, vehicles follow the currently assigned vehicle plans, can be boarded by customers, and can be charged. In the time-triggered fleet control phase, which will be described in more detail later, the operator can optimize its vehicle plans, e.g. by re-assigning users to other vehicles, repositioning vehicles in anticipation of future demand, or making new vehicle charging plans.

These core simulation modules can be extended by upper-level frameworks, such as a mobility broker~\cite{Engelhardt.2022} or an upper-level optimization problem to address the regulation of MoD services~\cite{Dandl.2021}. Contrarily, it is also possible to combine parts of these FleetPy standard simulation modules with other simulation packages. Most notably, FleetPy has been interfaced with mobiTopp\footnote{https://github.com/kit-ifv/mobitopp}~\cite{Wilkes.2021}, a state-of-the-art demand modeling software, to study the impact of ride-pooling with high-fidelity demand model based on empirically researched travel behavior~\cite{Wilkes.2021b}.

\subsection{Routing Modules and Vehicle Movements}
The network representation within FleetPy's routing modules takes an important role. On the one hand, it defines how vehicle movements are modeled and, on the other hand, it provides functionality for routing queries. Depending on the research question varying details in representing the network are desirable. Possible representations range from Euclidean planes, graph representations with constant travel times to graph representations with dynamic and/or stochastic travel times or even mesoscopic and microscopic traffic flow models used in commercial simulators. 

For the main use cases of FleetPy, the network is represented as a directed graph $G=(N,E)$ with nodes $N$ and edges $E$. An edge $e \in E$ is associated with a distance $d_e$ and travel time $t_e$. Positions in the network are represented as a tuple $(n_s, n_e, f)$. $n_s \in N$ and $n_e \in N$ correspond to start and end node of an edge. $f \in [0,1[$ reflects the fraction of edge distance that has been passed. If the position corresponds to a node itself, it is encoded either as $f=0$ or as not a value (None) for $n_e$ and $f$. A route is defined as a list of connected nodes. Routes, route travel times and distances are generally computed using Dijkstra and bi-directional Dijkstra algorithm. To solve the assignment problem for MoD services in an acceptable computation time, fast routing queries are essential because travel times between network positions are queried in a high frequency to check feasible routing combinations. Therefore, some speed-up techniques have been implemented to minimize the computational time for routing queries: 1) A C++ implementation of Dijkstra and bi-directional Dijkstra is available for considerable speed-up compared to the Python implementation. 2) In case the network size is small enough and travel times remain constant over a period of time, a module is available to exploit preprocessed travel time tables. 3) If the network size is too large for full preprocessing, it is possible to preprocess tables between the most important network nodes and fall back to implemented route computation algorithms for the rest of the nodes.

\subsection{Vehicles and Vehicle Route Legs}

Vehicles are modeled as explicit agents and interact with other modules in FleetPy: They receive tasks from their dedicated operators, move according to the routing modules, and interact with travelers for boarding and with charging operators for charging. Vehicles are described by two sets of attributes: vehicle-type dependent and dynamic attributes. Vehicle-type attributes such as maximum passenger capacity, range, battery capacity and cost attributes are fixed after initialization.

Contrarily, dynamic attributes describe the current status of the vehicle: position, set of on-board customers, state of charge (SOC) and a route if currently driving. Each vehicle performs a series of tasks defined by vehicle route legs. Route legs also describe the type of task, i.e. whether it is a driving, boarding or charging task. Accordingly, a route leg contains a route on the network (i.e. a sequence of nodes to pass) if it is a driving task, a list of boarding or alighting travelers if it is a boarding task or the assigned charging socket if it is a charging task. It contains a target position for the destination location if it is a driving task or the location where the task is performed for other tasks. Additionally, it keeps information regarding time constraints, for example the earliest start time or the minimum stop duration.

While being processed, information about the execution of the leg, i.e. start and end time, driven distance or information about the boarding or charging process is tracked and written to the output once the task is finished or terminated as described in Section \ref{subsec:input_output_data}. Furthermore, the progress of the execution of route legs is dynamically sent to the corresponding fleet control module.

\subsection{Demand Module}

The demand module handles all travelers (more specifically, a trip any traveler plans to make) during the simulation. At the beginning of a FleetPy simulation, travelers are loaded from an input file and initialized.

The decision-making behavior of travelers can be specified according to the use case. Possible behavior types include just selecting the first received offer, only accepting offers satisfying certain constraints, mode choice models (i.e. logit models) or even time-dependent choice behavior (for example a maximum response time for receiving an offer from the MoD operator). Depending on the specified traveler module, additional parameters like explicit time constraints or coefficients for choice models are initialized.

During the simulation, the demand module manages active travelers: New requests are revealed dynamically to all mobility operators. Generally, it is possible for travelers to request trips not only from multiple MoD operators but also alternative mode options like public transportation or private vehicles that can be represented in the simulation. The demand module thereby tracks travelers that did not request a trip yet, are waiting for a trip offer, or are currently traveling and is responsible to trigger writing traveler data to the output file.

\subsection{Infrastructure Module}
The infrastructure module models all tasks responsible for fleet maintenance. Static network locations for parking and/or charging can be initialized in this module. To model the interaction between infrastructure operators (e.g. charging operators) and MoD operators, multiple infrastructure operators can be initialized. Generally, FleetPy distinguishes between public infrastructure, which is accessible for all mobility services, and depot infrastructure, which is uniquely dedicated to a single MoD operator.

The module provides interfaces to query closest free parking spots and closest charging possibilities depending on a vehicle's position, its SOC and the charging power of the charging socket. The latter two parameters are used to estimate the required charging time and use it to find the empty time-slot at the nearest charging station. Besides, an explicit booking system is implemented to guarantee that each charging socket and parking space can only be occupied by one vehicle: Once a charging option is queried, it is referenced by an id specifying the charging slot and the expected duration for the charging process. After booking the charging option, the corresponding slot is reserved for the charging task and cannot be booked by other operators or vehicles. Additionally, the booking id is returned to the mobility operator to insert the information into the route leg of the corresponding vehicle to be charged.

\subsection{Fleet Control Module}
The goal of the fleet control module is to model the decision-making of an MoD operator and its interaction with other agents in FleetPy. Generally, a customer requests a trip from the operator and expects a feedback whether the request can be fulfilled and, if so, it also expects trip details on which it could base its choice (i.e. expected waiting time, travel time and/or fare). The operator also interacts with its fleet of vehicles by assigning schedules or tasks to pick up and drop off customers while -- by interaction with the Infrastructure Module -- charging processes are planned to guarantee continuous vehicle availability for the service.

In the fleet control module, different methods can be implemented to solve this dynamic and stochastic demand management and multi-vehicle routing problem. To represent this optimization problem, the Fleet Control Module makes use of three main objects: 1) Vehicle objects contain information about their current states like position, the number and list of on-board customers and state-of-charge (SOC). 2) \textit{PlanRequest} objects collect information about customers that requested a trip from the operator. This information includes request time, origin and destination position, group size and also (communicated or assumed) time constraints regarding pick-up and drop-off time windows. The time constraints can be either customer specific (communicated through the app) or MoD operator specific, i.e. the operator wants to ensure a certain level of service. 3) \textit{VehiclePlan} objects specify an ordered (hypothetical) task list for vehicles and therefore represent possible solution options of the underlying vehicle routing problem. Each task refers to a position with timing information (i.e. duration or earliest start time of the task) and a task specification (i.e. customers to board or alight or details regarding a charging process). Vehicle movements are planned by calculating routes in between the positions of the tasks by accessing the Routing Module. VehiclePlans are rated and compared by a customizable objective function which reflects the operator objective when assigning VehiclePlans to its vehicles. It is important to note that PlanRequest and VehiclePlan objects are not the same as Travelers and Vehicle Route Legs, respectively. The former only reflect information and hypothetical plans of the operator, while the latter define the actual actions. Thereby, the impacts of lack of information on the service operation can be modeled (i.e. stochastic travel times or unexpected user behavior). Once a VehiclePlan is chosen for assignment, it is converted to a list of Vehicle Route Legs and the current Vehicle Route Legs of the corresponding vehicle are replaced.

The approach in FleetPy to solve the dynamic and stochastic multi-vehicle routing problem is to divide the overall problem into sub-problems that are solved in corresponding sub-modules. In comparison to solution approaches like approximate dynamic programming \cite{powell2007approximate, AlKanj.2020}, this partitioning allows selecting modules with applicable solution algorithms for specific use cases and research questions. Especially for large scale problems with thousands of vehicles and customers, unified solution approaches are no longer applicable due to intractable computational time (e.g. see unified model-predictive control approach with 30 vehicles and 30 customers in \cite{R.Zhang.2016}). In the following the sub-modules are described on a high level.

\subsubsection{Immediate Response}
The response module is used to provide a customer feedback regarding service availability (triggered by \textit{user\_request} in Fig.~\ref{fig: subclasses flow chart}). This feedback (or service offer) includes attributes like expected pick-up time, travel time and/or fare and serves as input for the customer's decision-making. In the current implementation, an insertion heuristic is used to insert customer pick-up and drop-off in the fleet vehicles' currently assigned VehiclePlans. From the best plan according to the operator's objective, attributes for the service offer can be retrieved.

\subsubsection{Batch Optimization}
Within a selected set of time steps, the simulation module can trigger (\textit{time\_trigger} in Fig.~\ref{fig: subclasses flow chart}) global (re-)assignment optimization algorithms. The goal of the optimization is to treat all currently active requests in a batch and find the best possible solution for the corresponding static problem based on the current fleet state. In FleetPy a variant of the algorithm proposed by \citep{AlonsoMora.2017} is implemented for the ride-pooling use case. By exploiting time constraints for pick-up and maximum in-vehicle time, all feasible VehiclePlans for all vehicles serving a specific set of requests can be created. By solving an Integer Linear Problem (ILP) the best VehiclePlans are assigned to the vehicles. See \citep{Engelhardt.29.07.2020} for details.

\subsubsection{Repositioning}
Due to asymmetric pick-up and drop-off distributions of customers served by the MoD operator, vehicles tend to accumulate in regions where demand is lower than the flow of incoming vehicles. This often results in a degraded service rate or high average waiting times for customers. Pro-active repositioning algorithms can be deployed to anticipate future demand. If demand forecast with respect to a given zone system is available and provided to FleetPy, repositioning algorithms can be selected for simulation. Algorithms by \citep{Zhang.2016} and \citep{Syed.2021} are currently implemented.

\subsubsection{Reservation}
Some use cases require customers to book a trip in advance. Trips far in the future tend to heavily increase the computational time of the batch optimization algorithm because of a significant enlargement of the solution space. In FleetPy reservation requests are therefore separated from on-demand requests: A reserved trip thereby is included in a vehicle's VehiclePlan by solving an insertion heuristic, while leaving the request out of the batch optimization. Only when the pick-up time of this request approaches a rolling horizon, it is treated in the batch optimization.

\subsubsection{Charging}
The task of the charging module is to assign charging tasks to vehicles to prevent vehicles running out of charge. In case public or depot charging infrastructure is initialized, the charging module can book time slots for charging and assign them to vehicles. The currently implemented charging strategy assigns charging tasks at the closest charging station to vehicles once their SOC undercut a certain threshold.

\subsubsection{Dynamic Fleet Sizing}
The Dynamic Fleet Sizing module models time-dependent availability of vehicles, for example due to driver shifts or withholding of vehicles to balance demand and supply. Besides, maintenance can be conducted for these idle vehicles. In current implementations, it is either possible to define a time-dependent active vehicle distribution a priori, or a controller dynamically activates and deactivates vehicles to maintain a certain target utilization based on current demand (e.g. to determine a time-dependent required fleet size throughout the simulation period). Vehicles are deactivated by sending them to the next free depot and are no longer available for serving customers.

\subsubsection{Dynamic Pricing}
The Dynamic Pricing module sets fares dynamically during the simulation to model surge pricing. In the current implementation either a time-dependent approach or a fleet-utilization based pricing method can be used to scale fares.

\section{Reference FleetPy Showcase}

To demonstrate the flexibility and transferability of FleetPy, a case study for Manhattan, NYC is described in the following. Therefore, different models for simulation flow, customer behavior and network representation are shown and their impact on simulation results is evaluated.

\subsection{Input Data}

\begin{figure}
     \centering
     \begin{subfigure}[b]{\textwidth}
         \includegraphics[width=\textwidth]{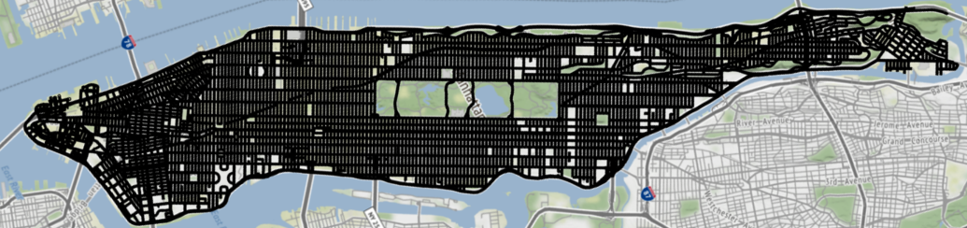}
        \caption{Applied street network of Manhattan (black).}
        \label{fig:map}
     \end{subfigure}
     \hfill
     \begin{subfigure}[b]{\textwidth}
         \includegraphics[width=\textwidth]{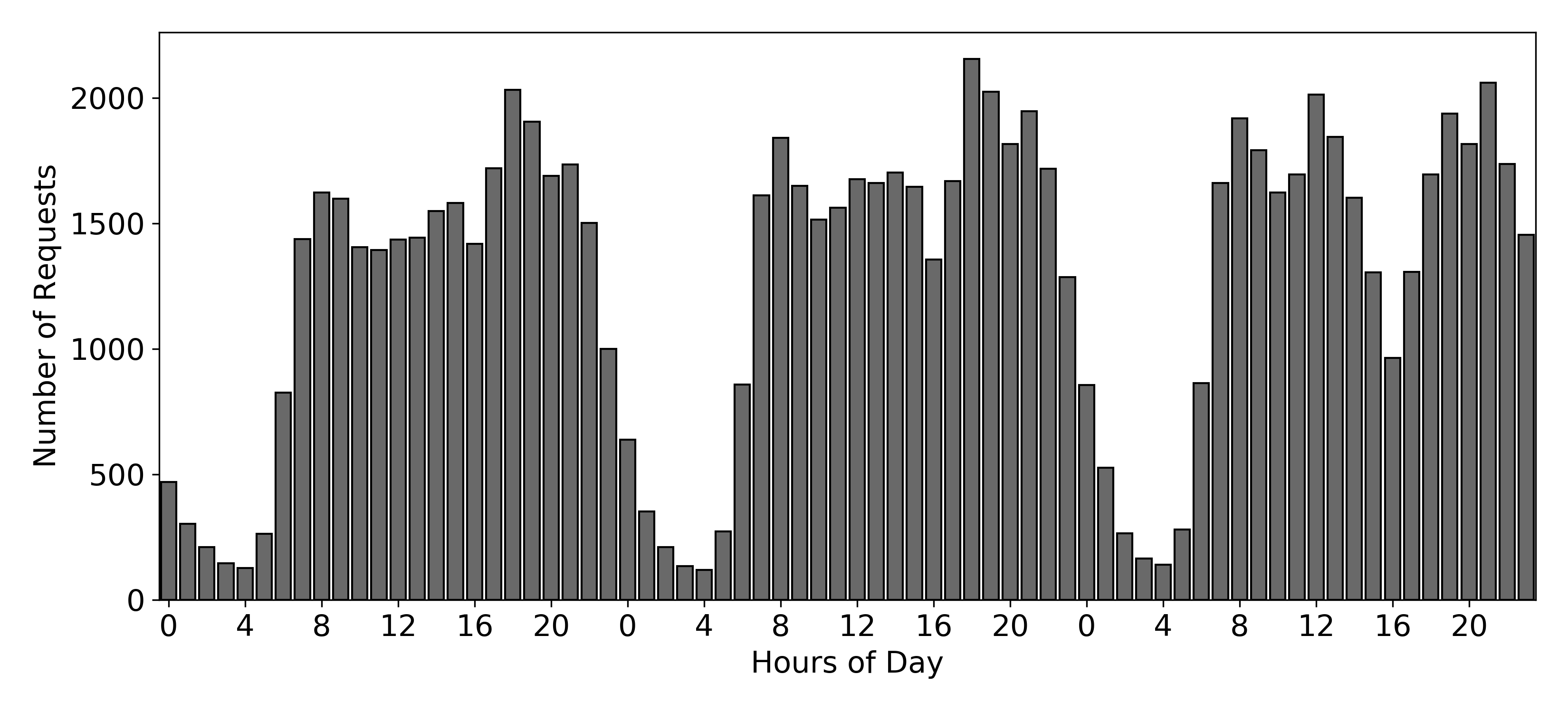}
        \caption{Number of requests from 6 June to 8 June 2016.}
        \label{fig:customers}
    \end{subfigure}
        \caption{Network and number of request per hour for the applied case study of Manhattan, NYC.}
        \label{fig:case_study}
\end{figure}

The publicly available taxi data set for NYC\footnote{https://www1.nyc.gov/site/tlc/about/tlc-trip-record-data.page} from 6 June to 8 June 2016 is used for this study. Besides the exact origin and destination coordinates, the data also include individual travel times and distances. This helps to create time-dependent link travel times with the assumption that historic trips took the shortest distance routes similar to \cite{Bertsimas.2019b}. The street network of Manhattan, NYC is extracted from OpenStreetMap (Fig.~\ref{fig:map}). Taxi trips starting and ending within Manhattan from this time period are used as demand for the simulated ride-pooling service: Origin and destination coordinates are matched onto closest network nodes while the start time of the taxi trip is used as a request time for the ride-pooling customer. To decrease computation time, only a 10\% subsample of all resulting trips are used to generate the MoD customers in the simulation (Fig.~\ref{fig:customers}).

The following setting is applied as base for all simulations if not explicitly mentioned otherwise: The service provider operates a fleet of 380 vehicles with capacity 4 and infinite range (no charging). The BOS-module is used for the simulation, i.e. customer requests receive trip-offers in batches. Also, the operator (re-)optimizes vehicle-customer assignment in batches every $60$s. Schedules have to fulfill a maximum waiting time constraint of $360$s and a maximum relative detour constraint of $40$\%. If a solution for a customer is found, the offer is always accepted by the customer, else the customer leaves the system unserved. It takes $30$s to board or alight a vehicle. Finally, the operator tries to assign schedules that fulfill all time and capacity constraints and serve as many customers as possible while minimizing the system time, i.e. the aggregated time needed to finish all schedules.

\subsection{Results}

\begin{figure}
    \centering
    \includegraphics[width=.6\textwidth]{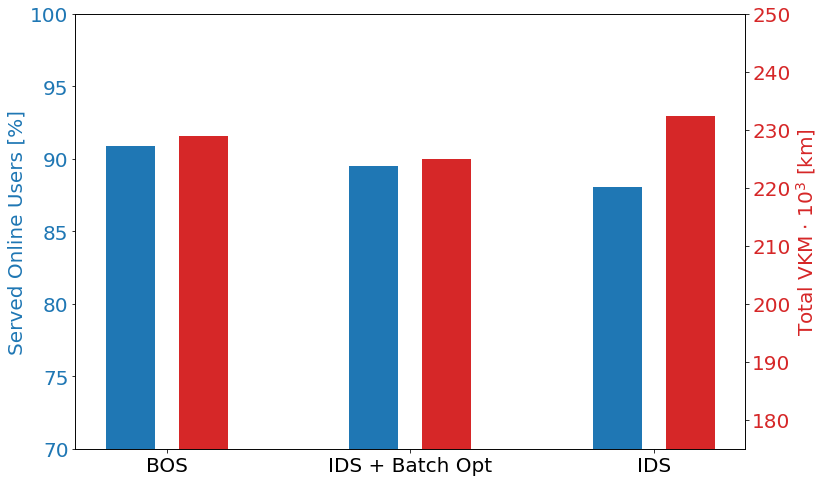}
    \caption{Served Users and Vehicle Kilometers traveled for different user interactions and modules}
    \label{fig:simflow_userInteraction}
\end{figure}

In the first example different simulation and assignment procedures are compared. Fig.~\ref{fig:simflow_userInteraction} shows the fraction of served customers and fleet vehicle kilometers (VKMs) traveled for three different scenarios: the described BOS framework of the base case, and two scenarios with IDS, where users get an offer and make their decision immediately after requesting a ride. While in the IDS framework the operator uses an insertion heuristic to create schedules and offers, a periodic re-optimization is used in IDS with batch opt. Fig.~\ref{fig:simflow_userInteraction} shows that the IDS-scenario serves the least customers while traveling the most. A direct comparison of IDS with batch opt shows that the periodic re-optimization can find faster and shorter routes to serve the customers and increase vehicle availability. Served customers and also VKM increases slightly in the BOS-scenario compared to IDS with batch opt. While the IDS framework forces the operator to immediately make a decision for requests to accept or reject, in BOS it can decide for all new requests together in batch increasing the solution space of the optimization. Therefore the primary objective of serving customers can be improved.

\begin{figure}
     \centering
     \begin{subfigure}[b]{0.49\textwidth}
         \includegraphics[width=\textwidth]{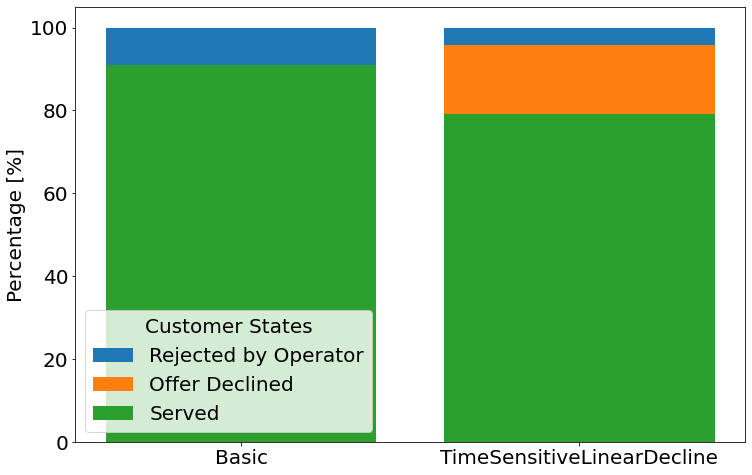}
        \caption{Customer states after simulation}
        \label{fig:userModel_requests}
     \end{subfigure}
     \hfill
     \begin{subfigure}[b]{0.49\textwidth}
         \includegraphics[width=\textwidth]{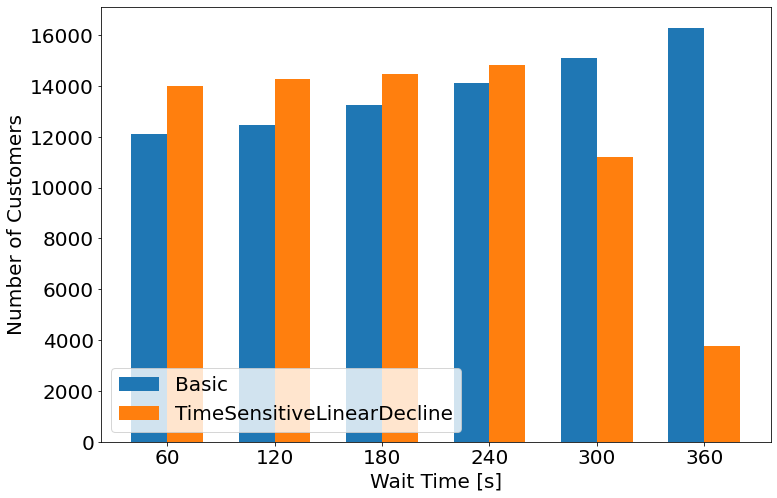}
        \caption{Waiting time distribution of served customers}
        \label{fig:userModel_wait_time}
    \end{subfigure}
        \caption{Comparison of BasicRequest and TimeSensitiveLinearDeclineRequest.}
        \label{fig:user_model}
\end{figure}

Next, Fig.~\ref{fig:user_model} compares the effect of two different behavioral models for MoD-customers. BasicRequest corresponds to the base case: The customers always accept an offer (that satisfies the time constraints) from the operator. The TimeSensitiveLinearDeclineRequest on the other hand can reject an offer based on the offered waiting time. Here, the offer is always accepted if the offered waiting time is lower than $240$s, from $240$s to $360$s the acceptance probability decreases linearly, and waiting times longer than $360$s are never accepted. As expected, Fig.~\ref{fig:userModel_requests} shows that less customers are served in the latter scenario, because customers start declining the service. On the other hand, the fraction of customers that do not receive an offer at all decreases, because vehicles serve less customers and, thereby, have a higher probability of being available. Fig.~\ref{fig:userModel_wait_time} shows the waiting time distribution of served customers: In the scenario with BasicRequests, the operator uses the whole range of waiting time to make assignments and a trend for more customers with higher waiting times is observable. Without adaption of the fleet control, it can be expected that the offered waiting times follow the trend of the scenario with the BasicRequests; however, the fraction of TimeSensitiveLinearDeclineRequests with waiting times higher than $240$s is strongly reduced because offers are declined.

\begin{figure}
    \centering
    \includegraphics[width=1\textwidth]{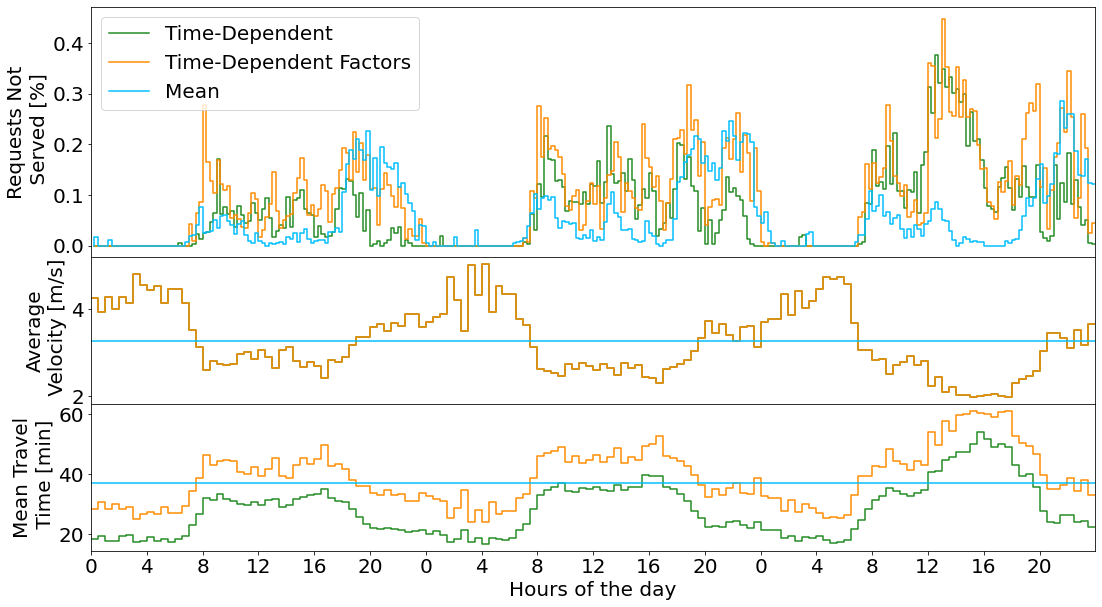}
    \caption{Top: Percentage of unserved customers and average edge velocity of the system over the whole simulation duration. The unserved requests are presented in bins over 15 minutes. Middle:  Average network velocity $v_{avg}(t)$ is presented as bins over 30 minutes. The average velocities of "Time-Dependent" and "Time-Dependent Factors" are lying on top if each other. Bottom: Average travel time of fastest routes between 250 randomly drawn nodes in the network.}
    \label{fig:NWTT}
\end{figure}

Different network representations can have significant effects on the performance of MoD-services. To demonstrate these effects, Fig.~\ref{fig:NWTT} shows the effect on unserved customers over the course of the simulation. Three different network representations are compared: "Time-Dependent" refers to the base representation, i.e. time-dependent and edge-specific travel times. "Time-Dependent Factors" refers to a variant which is useful when only average network velocities are known: All edge travel times are based on their free flow travel times and scaled by the same factor. The factors are determined to fit the average network velocity $v_{avg}(t)$ within a certain time interval, defined by

  \begin{align}
    v_{avg}(t) = \frac{\sum_{i \in E} d_e}{\sum_{i \in E} t_e(t)} ~.
\label{eq:tred}
    \end{align}

"Mean" corresponds to the same method as "Time-Dependent Factors" but only one single factor is used for the whole simulation representing the time-average of the dynamic factors. Fig.~\ref{fig:NWTT} compares three KPIs related to the different network representations: On top, unserved requests due to high demand are shown. Depending on the time and network representation, most requests are declined during high demand times in the morning, midday and evening. With faster travel times between pick-ups and drop-offs, fleet vehicles can serve more customers in less time resulting in less requests not served. In the middle, the average network velocities are shown. By design, "Time-Dependent" and "Time-Dependent Factors" share the same average velocity $v_{avg}(t)$. When comparing the trends of average velocity and requests not served of the "Mean" and "Time-Dependent Factors" representation, it can be seen that the more requests can be served if the velocity is higher. Nevertheless, due to network inhomogeneities added in the "Time-Dependent" representation, this trend can not always be observed. To show the effect of increased inhomogeneity the average fastest travel time between 250 randomly drawn nodes within the network is plotted on the bottom. Even though the mean network travel times are the same for "Time-Dependent" and "Time-Dependent Factors", the average travel times of the fastest routes are decreased. This effect explains especially the trend in times from 9pm to 12pm that less unserved customers are observed for the "Time-Dependent" even though the same network velocity as in "Time-Dependent Factors" is applied.

\begin{figure}
    \centering
    \includegraphics[width=.6\textwidth]{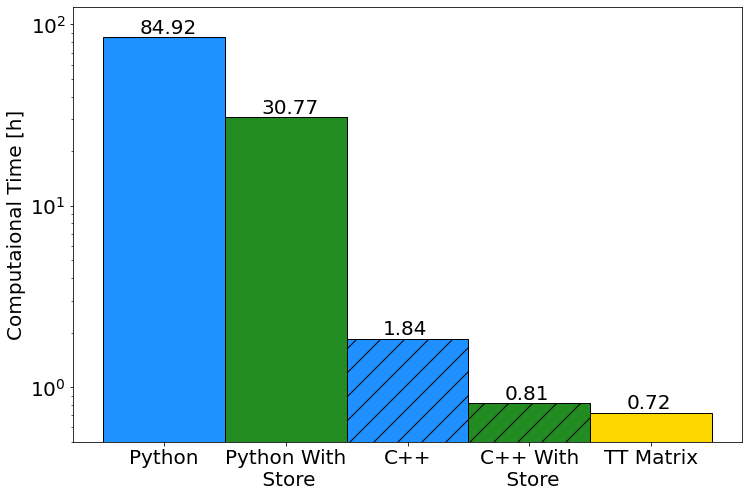}
    \caption{Comparison of the sum of the computational time of user request and request batch between different network representations. Computational time was collected for simulations of nine hour periods. The y-axis is presented on a logarithmic scale.}
    \label{fig:NWRepresentation}
\end{figure}

Lastly, Fig.~\ref{fig:NWRepresentation} shows the computation time of different routing modules, which mainly differ in their implementations of querying network travel times and routes. The "TT Matrix" uses a fully preprocessed travel time table to look up travel times in the network and builds routes according to~\cite{Dandl.2020}. The other modules implement Dijkstra and bi-directional Dijkstra algorithms in "Python" and "C++". In the "C++" module, the routing algorithms are implemented in the C++ programming language and included in the Python code of FleetPy using the CPython library. The supplement "With Store" means that once travel information between an origin-destination pair is computed, the result is stored and directly returned if the same routing information is queried again. This storage is emptied if network travel times are updated. It can be observed in Fig.~\ref{fig:NWRepresentation} that the "TT Matrix" module is fastest because only table look-up is required for calculating many-to-many travel times and distances and routes can be created slightly more efficient. However, actual routes are rarely queried compared to travel times and distances only. Nevertheless, if the network is too big, preprocessing the travel time tables will no longer be possible as too much storage is consumed. As a trade-off, storing only already computed routing information provides a speed-up of 2.79 for the Python implementation. Additionally, using an implementation within C++ gives an enormous speed-up of 104.8 compared to the Python implementation of Dijkstra. Overall the evaluation of the total computational time shows that choosing the right routing algorithms is critical for being able to run many simulations because routing queries consume the majority of the computational time.

\section{Conclusion}
\subsection{Summary}

In this paper, the agent-based MoD fleet simulation framework FleetPy is presented. With the advancement in digitalization, MoD services and sharing economy are blooming, bringing new options beyond traditional private and public transportation modes, and already occupying significant urban mobility shares in many cities. With current mobility trends and challenges in urban and non-urban areas MoD services can increase vehicle utilization (and therefore reduce space required for parking), decrease vehicle kilometers traveled by sharing rides, and increase accessibility to public transportation if implemented properly. Simulations can be used to study the impact of different MoD implementation strategies and give recommendations to stakeholders. With this background, FleetPy is conceptualized to study both the economic and social influence of MoD services.

Based on a literature review of existing simulation frameworks to model MoD services, following strengths of FleetPy can be determined:
\begin{itemize}
    \item Realistic customer-operator interaction: Interactions between users and operator(s) can be modeled by a request-offer-customer decision process based on real-time information. This decision process not only includes mode choices but also dynamic behavior.
    \item Flexibility: Multiple (MoD) providers can be integrated to study upper-level effects on the transportation system by including broker or regulatory measures.
    \item Modularity and Transferability: Separation of different tasks into modules allows easy reusing of already implemented modules and selection of the suitable modeling detail while new control algorithms can be implemented and tested by extending the current modules.
\end{itemize}

The core simulation module of FleetPy initializes the simulation based on the provided parameters and input data. The simulation flow is specified here and triggers vehicle and network state updates, optimization and other control procedures for fleet control and especially defines how customers and operators interact. Two different implementations are presented: IDS models an immediate customer-operator interaction, while in BOS the customers request trips from fleet operators in batches and do not expect an immediate reply. Both implementations allow modeling a complex and dynamic behavior of customer decision processes.

Besides the core simulation module, other transferable modules in FleetPy comprise of network and routing, demand and travelers, infrastructure and fleet control with corresponding sub-modules. A strict separation of traveler and vehicle route legs from planning classes in the fleet control module allows studying the impact of incomplete knowledge of an operator. Fleet control sub-modules provide control algorithms solving different sub-tasks of the overall fleet control problem, such as customer-vehicle assignments, repositioning, reservation, charging, dynamic fleet sizing and dynamic pricing.

Finally, some example studies are presented that show the impact of using different FleetPy modules. The simulation flow (IDS or BOS), customer interaction and operator optimization directly influence the number of served customers and fleet VKM. Finally, the computation time is shown to be affected by different network models and routing algorithms.

\subsection{Future Work}
There are multiple aspects we plan to investigate with FleetPy in the future. Firstly, the impact of new MoD service concepts will be explored. For example, the underutilized vehicles in times of low demand can be used for parcel transport~\cite{Fehn.10.05.2022}. Simulation can be used to evaluate possible benefits and whether this integration is a business case for operators. Moreover, the FleetPy control module will be connected to a front-end for a real-world application. This will highlight the high level of detail of the simulation framework and the usability for operators to develop and evaluate realistic fleet control strategies with FleetPy. Further work is planned on interfaces with MATSim and AIMSUN to have more detailed traffic representations and evaluations. Hereby, the approach shown in \citep{Dandl.2017} will be updated. Finally, implementations of high-performance routing algorithms, such as customizable contraction hierarchies~\cite{Dibbelt.2016}, and further developments of mesoscopic traffic flow representations with macroscopic dynamics are on the list of planned updates.


\section*{Author Contributions}
The authors confirm contribution to the paper as follows: study conception and design: Roman Engelhardt, Florian Dandl, Klaus Bogenberger; development of methodology: Roman Engelhardt, Florian Dandl, Arslan-Ali Syed, Fabian Fehn; software implementation: Roman Engelhardt, Florian Dandl, Arslan-Ali Syed, Yunfei Zhang, Fynn Wolf; draft manuscript preparation: Roman Engelhardt, Florian Dandl, Arslan-Ali Syed, Yunfei Zhang, Fabian Fehn, Fynn Wolf. All authors reviewed the results and approved the final version of the manuscript.

\bibliographystyle{plainnat}
\bibliography{references}

\begin{thebibliography}{39}
\providecommand{\natexlab}[1]{#1}
\providecommand{\url}[1]{\texttt{#1}}
\expandafter\ifx\csname urlstyle\endcsname\relax
  \providecommand{\doi}[1]{doi: #1}\else
  \providecommand{\doi}{doi: \begingroup \urlstyle{rm}\Url}\fi

\bibitem[Acheampong et~al.(2020)Acheampong, Siiba, Okyere, and
  Tuffour]{acheampong2020mobility}
Ransford~A Acheampong, Alhassan Siiba, Dennis~K Okyere, and Justice~P Tuffour.
\newblock Mobility-on-demand: An empirical study of internet-based ride-hailing
  adoption factors, travel characteristics and mode substitution effects.
\newblock \emph{Transportation Research Part C: Emerging Technologies},
  115:\penalty0 102638, 2020.

\bibitem[Adnan et~al.(2016)Adnan, Pereira, Lima~Azevedo, Basak, Lovric, Raveau,
  Zhu, Ferreira, Zegras, and Ben-Akiva]{Adnan2016}
Muhammad Adnan, Francisco Pereira, Carlos Lima~Azevedo, Kakali Basak, Milan
  Lovric, Sebastián Raveau, Yi~Zhu, Joseph Ferreira, Chris Zegras, and Moshe
  Ben-Akiva.
\newblock Simmobility: A multi-scale integrated agent-based simulation
  platform.
\newblock 01 2016.

\bibitem[{Aimsun SLU}(2022)]{AimsunRide2022}
{Aimsun SLU}.
\newblock Aimsun ride, 2022.
\newblock URL \url{https://www.aimsun.com/aimsun-ride-research-program/}.

\bibitem[Al-Kanj et~al.(2020)Al-Kanj, Nascimento, and Powell]{AlKanj.2020}
Lina Al-Kanj, Juliana Nascimento, and Warren~B. Powell.
\newblock Approximate dynamic programming for planning a ride-hailing system
  using autonomous fleets of electric vehicles.
\newblock \emph{European Journal of Operational Research}, 284\penalty0
  (3):\penalty0 1088--1106, 2020.
\newblock ISSN 03772217.
\newblock \doi{10.1016/j.ejor.2020.01.033}.

\bibitem[Alonso-Mora et~al.(2017)Alonso-Mora, {S. Samaranayake}, {A. Wallar},
  {E. Frazzoli}, and {D. Rus}]{AlonsoMora.2017}
J.~Alonso-Mora, {S. Samaranayake}, {A. Wallar}, {E. Frazzoli}, and {D. Rus}.
\newblock On-demand high-capacity ride-sharing via dynamic trip-vehicle
  assignment.
\newblock \emph{Proceedings of the National Academy of Sciences}, 114\penalty0
  (3):\penalty0 462--467, 2017.

\bibitem[Auld et~al.(2016)Auld, Hope, Ley, Sokolov, Xu, and Zhang]{Auld2016}
Joshua Auld, Michael Hope, Hubert Ley, Vadim Sokolov, Bo~Xu, and Kuilin Zhang.
\newblock Polaris: Agent-based modeling framework development and
  implementation for integrated travel demand and network and operations
  simulations.
\newblock \emph{Transportation Research Part C: Emerging Technologies},
  64:\penalty0 101--116, 2016.
\newblock ISSN 0968-090X.
\newblock \doi{https://doi.org/10.1016/j.trc.2015.07.017}.
\newblock URL
  \url{https://www.sciencedirect.com/science/article/pii/S0968090X15002703}.

\bibitem[Bertsimas et~al.(2019)Bertsimas, Delarue, Jaillet, and
  Martin]{Bertsimas.2019b}
Dimitris Bertsimas, Arthur Delarue, Patrick Jaillet, and S{\'e}bastien Martin.
\newblock Travel time estimation in the age of big data.
\newblock \emph{Operations Research}, 2019.
\newblock ISSN 0030-364X.
\newblock \doi{10.1287/opre.2018.1784}.

\bibitem[B{\"o}sch et~al.(2018)B{\"o}sch, Becker, Becker, and
  Axhausen]{Bosch.2018}
Patrick~M. B{\"o}sch, Felix Becker, Henrik Becker, and Kay~W. Axhausen.
\newblock Cost-based analysis of autonomous mobility services.
\newblock \emph{Transport Policy}, 64\penalty0 (11):\penalty0 76--91, 2018.
\newblock ISSN 0967070X.
\newblock \doi{10.1016/j.tranpol.2017.09.005}.

\bibitem[Dandl and Bogenberger(2019)]{Dandl.2019b}
Florian Dandl and Klaus Bogenberger.
\newblock Comparing future autonomous electric taxis with an existing
  free-floating carsharing system.
\newblock \emph{IEEE Transactions on Intelligent Transportation Systems},
  20\penalty0 (6):\penalty0 2037--2047, 2019.
\newblock ISSN 1524-9050.
\newblock \doi{10.1109/TITS.2018.2857208}.

\bibitem[Dandl et~al.(2017)Dandl, Bracher, and Bogenberger]{Dandl.2017}
Florian Dandl, Benedikt Bracher, and Klaus Bogenberger.
\newblock Microsimulation of an autonomous taxi-system in munich.
\newblock In \emph{2017 5th IEEE International Conference on Models and
  Technologies for Intelligent Transportation Systems (MT-ITS)}, pages
  833--838, 2017.
\newblock \doi{10.1109/MTITS.2017.8005628}.

\bibitem[Dandl et~al.(2020)Dandl, Tilg, Rostami-Shahrbabaki, and
  Bogenberger]{Dandl.2020}
Florian Dandl, Gabriel Tilg, Majid Rostami-Shahrbabaki, and Klaus Bogenberger.
\newblock Network fundamental diagram based routing of vehicle fleets in
  dynamic traffic simulations.
\newblock In \emph{2020 IEEE 23rd International Conference on Intelligent
  Transportation Systems (ITSC)}, pages 1--8. IEEE, 2020.
\newblock ISBN 978-1-7281-4149-7.
\newblock \doi{10.1109/ITSC45102.2020.9294204}.

\bibitem[Dandl et~al.(2021)Dandl, Engelhardt, Hyland, Tilg, Bogenberger, and
  Mahmassani]{Dandl.2021}
Florian Dandl, Roman Engelhardt, Michael Hyland, Gabriel Tilg, Klaus
  Bogenberger, and Hani~S. Mahmassani.
\newblock Regulating mobility-on-demand services: Tri-level model and bayesian
  optimization solution approach.
\newblock \emph{Transportation Research Part C: Emerging Technologies},
  125:\penalty0 103075, 2021.
\newblock ISSN 0968-090X.
\newblock \doi{10.1016/j.trc.2021.103075}.

\bibitem[Dibbelt et~al.(2016)Dibbelt, Strasser, and Wagner]{Dibbelt.2016}
Julian Dibbelt, Ben Strasser, and Dorothea Wagner.
\newblock Customizable contraction hierarchies.
\newblock \emph{Journal of Experimental Algorithmics}, 21\penalty0
  (1):\penalty0 1--49, 2016.
\newblock ISSN 10846654.
\newblock \doi{10.1145/2886843}.

\bibitem[Engelhardt and Bogenberger(2021)]{Engelhardt.2021}
Roman Engelhardt and Klaus Bogenberger.
\newblock Benefits of flexible boarding locations in on-demand ride-pooling
  systems.
\newblock In \emph{2021 7th International Conference on Models and Technologies
  for Intelligent Transportation Systems (MT-ITS)}, pages 1--6, 2021.
\newblock \doi{10.1109/MT-ITS49943.2021.9529284}.

\bibitem[Engelhardt et~al.(2019)Engelhardt, Dandl, Bilali, and
  Bogenberger]{Engelhardt.2019}
Roman Engelhardt, Florian Dandl, Aledia Bilali, and Klaus Bogenberger.
\newblock Quantifying the benefits of autonomous on-demand ride-pooling: A
  simulation study for munich, germany.
\newblock In \emph{2019 22nd IEEE Intelligent Transportation Systems Conference
  (ITSC)}, pages 2992--2997. IEEE, 2019.
\newblock ISBN 978-1-5386-7024-8.
\newblock \doi{10.1109/ITSC.2019.8916955}.

\bibitem[Engelhardt et~al.(2020)Engelhardt, Dandl, and
  Bogenberger]{Engelhardt.29.07.2020}
Roman Engelhardt, Florian Dandl, and Klaus Bogenberger.
\newblock Speed-up heuristic for an on-demand ride-pooling algorithm, 2020.
\newblock URL \url{https://arxiv.org/pdf/2007.14877}.

\bibitem[Engelhardt et~al.(2022)Engelhardt, Malcolm, Dandl, and
  Bogenberger]{Engelhardt.2022}
Roman Engelhardt, Patrick Malcolm, Florian Dandl, and Klaus Bogenberger.
\newblock Competition and cooperation of autonomous ridepooling services:
  Game-based simulation of a broker concept.
\newblock \emph{Frontiers in Future Transportation}, 3, 2022.
\newblock ISSN 2673-5210.
\newblock \doi{10.3389/ffutr.2022.915219}.

\bibitem[{European Environment Agency}(2022)]{EEA2022}
{European Environment Agency}.
\newblock Atmospheric greenhouse gas concentrations, 2022.
\newblock URL
  \url{https://www.eea.europa.eu/ims/atmospheric-greenhouse-gas-concentrations}.

\bibitem[Fehn et~al.(2022)Fehn, Engelhardt, Dandl, Bogenberger, and
  Busch]{Fehn.10.05.2022}
Fabian Fehn, Roman Engelhardt, Florian Dandl, Klaus Bogenberger, and Fritz
  Busch.
\newblock Integrating parcel deliveries into a ride-pooling service -- an
  agent-based simulation study, 2022.
\newblock URL \url{https://arxiv.org/pdf/2205.04718}.

\bibitem[Heinrichs et~al.(2013)]{heinrichs2013sharing}
Harald Heinrichs et~al.
\newblock Sharing economy: a potential new pathway to sustainability.
\newblock \emph{GAIA-Ecological Perspectives for Science and Society},
  22\penalty0 (4):\penalty0 228--231, 2013.

\bibitem[Horni et~al.(2016)Horni, Nagel, and Axhausen]{Horni2016}
Andreas Horni, Kai Nagel, and Kay Axhausen, editors.
\newblock \emph{Multi-Agent Transport Simulation MATSim}.
\newblock Ubiquity Press, London, Aug 2016.
\newblock ISBN 978-1-909188-75-4, 978-1-909188-76-1, 978-1-909188-77-8,
  978-1-909188-78-5.
\newblock \doi{10.5334/baw}.

\bibitem[Huang et~al.(2022)Huang, Cui, Zhang, Tong, Shi, and Liu]{Huang2022}
Jiangyan Huang, Youkai Cui, Lele Zhang, Weiping Tong, Yunyang Shi, and Zhiyuan
  Liu.
\newblock An overview of agent-based models for transport simulation and
  analysis.
\newblock \emph{Journal of Advanced Transportation}, 2022:\penalty0 17, 2022.
\newblock \doi{https://doi.org/10.1155/2022/1252534}.
\newblock URL \url{https://www.hindawi.com/journals/jat/2022/1252534/}.
\newblock Article ID 1252534.

\bibitem[Kucharski and Cats(2022)]{Kucharski2022}
Rafał Kucharski and Oded Cats.
\newblock Simulating two-sided mobility platforms with maassim.
\newblock \emph{PLOS ONE}, 17\penalty0 (6):\penalty0 1--20, 06 2022.
\newblock \doi{10.1371/journal.pone.0269682}.
\newblock URL \url{https://doi.org/10.1371/journal.pone.0269682}.

\bibitem[Kuhnimhof and Eisenmann(2021)]{kuhnimhof2021mobility}
Tobias Kuhnimhof and Christine Eisenmann.
\newblock Mobility-on-demand pricing versus private vehicle tco: how cost
  structures hinder the dethroning of the car.
\newblock \emph{Transportation}, pages 1--25, 2021.

\bibitem[Nguyen et~al.(2021)Nguyen, Powers, Urquhart, Farrenkopf, and
  Guckert]{Nguyen2021}
Johannes Nguyen, Simon~T. Powers, Neil Urquhart, Thomas Farrenkopf, and Michael
  Guckert.
\newblock An overview of agent-based traffic simulators.
\newblock \emph{Transportation Research Interdisciplinary Perspectives},
  12:\penalty0 100486, 2021.
\newblock ISSN 2590-1982.
\newblock \doi{https://doi.org/10.1016/j.trip.2021.100486}.
\newblock URL
  \url{https://www.sciencedirect.com/science/article/pii/S2590198221001913}.

\bibitem[Powell(2007)]{powell2007approximate}
Warren~B Powell.
\newblock \emph{Approximate Dynamic Programming: Solving the curses of
  dimensionality}, volume 703.
\newblock John Wiley \& Sons, 2007.

\bibitem[{PTV Group}(2022)]{PTVMaaSModeller2022}
{PTV Group}.
\newblock Ptv maas modeller, 2022.
\newblock URL \url{https://www.ptvgroup.com/de/mobilitynext/}.

\bibitem[{R. Zhang} et~al.(2016){R. Zhang}, {F. Rossi}, and {M.
  Pavone}]{R.Zhang.2016}
{R. Zhang}, {F. Rossi}, and {M. Pavone}.
\newblock Model predictive control of autonomous mobility-on-demand systems.
\newblock In \emph{2016 IEEE International Conference on Robotics and
  Automation (ICRA)}, pages 1382--1389, 2016.
\newblock \doi{10.1109/ICRA.2016.7487272}.

\bibitem[Rodrigue(2020)]{Rodrigue2020}
Jean-Paul Rodrigue.
\newblock \emph{The Geography of Transport Systems (5th ed.)}.
\newblock Routledge, London, 2020.
\newblock \doi{https://doi.org/10.4324/9780429346323}.

\bibitem[Ruch et~al.(2018)Ruch, Hörl, and Frazzoli]{Ruch2018}
Claudio Ruch, Sebastian Hörl, and Emilio Frazzoli.
\newblock Amodeus, a simulation-based testbed for autonomous mobility-on-demand
  systems.
\newblock In \emph{2018 21st International Conference on Intelligent
  Transportation Systems (ITSC)}, pages 3639--3644, 2018.
\newblock \doi{10.1109/ITSC.2018.8569961}.

\bibitem[Schaller(2021)]{schaller2021can}
Bruce Schaller.
\newblock Can sharing a ride make for less traffic? evidence from uber and lyft
  and implications for cities.
\newblock \emph{Transport policy}, 102:\penalty0 1--10, 2021.

\bibitem[Soares et~al.(2014)Soares, Kokkinogenis, Macedo, and
  Rossetti]{Soares2014}
Guilherme Soares, Zafeiris Kokkinogenis, José Macedo, and Rosaldo Rossetti.
\newblock Agent-based traffic simulation using sumo and jade: An integrated
  platform for artificial transportation systems.
\newblock pages 44--61, 11 2014.
\newblock \doi{10.1007/978-3-662-45079-6_4}.

\bibitem[Standing et~al.(2019)Standing, Standing, and
  Biermann]{standing2019implications}
Craig Standing, Susan Standing, and Sharon Biermann.
\newblock The implications of the sharing economy for transport.
\newblock \emph{Transport Reviews}, 39\penalty0 (2):\penalty0 226--242, 2019.

\bibitem[Syed et~al.(2021)Syed, Dandl, Kaltenh{\"a}user, and
  Bogenberger]{Syed.2021}
Arslan~Ali Syed, Florian Dandl, Bernd Kaltenh{\"a}user, and Klaus Bogenberger.
\newblock Density based distribution model for repositioning strategies of ride
  hailing services.
\newblock \emph{Frontiers in Future Transportation}, 2:\penalty0 13, 2021.
\newblock ISSN 2673-5210.
\newblock \doi{10.3389/ffutr.2021.681451}.
\newblock URL
  \url{https://www.frontiersin.org/articles/10.3389/ffutr.2021.681451/full#B6}.

\bibitem[Tirachini(2020)]{tirachini2020ride}
Alejandro Tirachini.
\newblock Ride-hailing, travel behaviour and sustainable mobility: an
  international review.
\newblock \emph{Transportation}, 47\penalty0 (4):\penalty0 2011--2047, 2020.

\bibitem[{United Nations Department of Economic and Social
  Affairs}(2022)]{UN2022}
{United Nations Department of Economic and Social Affairs}.
\newblock 68\% of the world population projected to live in urban areas by
  2050, 2022.
\newblock URL
  \url{https://www.un.org/development/desa/en/news/population/2018-revision-of-world-urbanization-prospects.html}.

\bibitem[Wilkes et~al.(2021{\natexlab{a}})Wilkes, Engelhardt, Briem, Dandl,
  Vortisch, Bogenberger, and Kagerbauer]{Wilkes.2021}
Gabriel Wilkes, Roman Engelhardt, Lars Briem, Florian Dandl, Peter Vortisch,
  Klaus Bogenberger, and Martin Kagerbauer.
\newblock Self-regulating demand and supply equilibrium in joint simulation of
  travel demand and a ride-pooling service.
\newblock \emph{Transportation Research Record: Journal of the Transportation
  Research Board}, 2675\penalty0 (8):\penalty0 226--239, 2021{\natexlab{a}}.
\newblock ISSN 2169-4052.
\newblock \doi{10.1177/0361198121997140}.

\bibitem[Wilkes et~al.(2021{\natexlab{b}})Wilkes, Engelhardt, Kostorz, Dandl,
  Zwick, Fraedrich, and Kagerbauer]{Wilkes.2021b}
Gabriel Wilkes, Roman Engelhardt, Nadine Kostorz, Florian Dandl, Felix Zwick,
  Eva Fraedrich, and Martin Kagerbauer.
\newblock Assessing the effects of ride-pooling - a case study of moia in
  hamburg.
\newblock In \emph{ITS World Congress}, 2021{\natexlab{b}}.

\bibitem[Zhang and Pavone(2016)]{Zhang.2016}
Rick Zhang and Marco Pavone.
\newblock Control of robotic mobility-on-demand systems: A queueing-theoretical
  perspective.
\newblock \emph{The International Journal of Robotics Research}, 35\penalty0
  (1-3):\penalty0 186--203, 2016.
\newblock \doi{10.1177/0278364915581863}.

\end{thebibliography}

\end{document}